\documentclass[
,final            % use final for the camera ready runs
  ]
  {aipproc}
\usepackage{epsfig}
\layoutstyle{6x9}
\begin{document}

\title{The brain near the edge\footnote{Contribution to the 9th Granada  Seminar in Computational and Statistical Physics. Computational and Mathematical Modelling of Cooperative Behavior in Neural Systems, (2006). University of Granada, Spain. AIP Proceedings (2006, in press)}}


\author{Dante R. Chialvo }{address={Department of Physiology, Feinberg Medical
School, Northwestern University, 303 East Chicago Ave. Chicago, IL
60611, USA} }

\begin{abstract}
When viewed at a certain coarse grain, the brain seems a
relatively small dynamical system composed by a few dozen
interacting areas, performing a number of stereotypical behaviors.
It is known that, even relatively small dynamical systems can
reliably generate robust and flexible behavior if they are possed
near a second order phase transition, because of the abundance of
metastable states at the critical point. The approach pursued here
assumes that some of the most fundamental properties of the
functioning brain are possible because it is spontaneously possed
at the border of such instability. In this notes we review the
motivation, the arguments and recent results as well as the
implications of this view of the functioning brain.
\end{abstract}

\maketitle

\section{Introduction}
Each year, there are several hundredths of fascinating discoveries
of isolated aspects of brain physiology. At the same time, only a
handful or reports discusses the reverse process: how the
knowledge of isolated pieces can be integrated to explain how the
brain works. This is, of course, a well known, hard to tackle
challenge which, however, is particularly suitable to a physicist
because of it inherent familiarity with ideas of universality and
unification. For a newcomer the first concern would be if it is
possible to approach the problem of brain function without
inventing a new theoretical framework. In other words, is it
possible to gain any insight about relevant brain problems by
deliberately ignoring -at least for the moment- the soft aspects
of brain's condensed matter?

The approach pursued here assumes that the most fundamental
properties of the functioning brain are possible because is
spontaneously possed at the border of an instability. Indeed the
proposal is that these fascinating properties have no extra cost
as they are generic for this state. From this viewpoint, all human
behaviors, including thoughts, undirected or goal oriented
actions, or simply any state of mind, are the outcome of a
dynamical system -the brain- at or near a critical state. The main
point is that, as in thermodynamics systems at the critical point,
it is only at this state that the largest behavioral repertoire
can be attained by the smallest number of degrees of freedom.
\emph{Behavioral repertoire} means the set of actions useful for
the survival of the brain and \emph{degrees of freedom} means the
number of (loosely defined) specialized brain areas engaged in
generating such actions. By looking at the problem from this angle
a number of ideas and results from statistical physics can be used
to guide work towards the ultimate goal of understanding how the
brain works, without inventing anything new.

This article is dedicated to discuss the basis and the
implications of this proposition. The paper is organized as
follows. The second section introduces the main motivation, which
is routed in concepts borrowed from complex systems. For
completeness we also summarize here some obvious connections with
very well known facts from statistical physics. The third section
discuss the specific rationale and the next section enumerates
evidence that seems to support the usefulness of this approach to
brain function. The paper closes with a short discussion of
implications.

\section{Fundamental Laws for the Collective}

It is well known that almost all interesting macroscopic phenomena
in nature, from gravity to photosynthesis, from superconductivity
to muscle contraction are product of an underlying
\emph{collective} phenomena. In this sense, science is  the never
ending process of explaining macroscopic phenomena observed at one
level from fundamental laws uncovered at another level.
Neuroscience not being an exception, must explain human behavior,
i.e., \emph{what we see} in terms of the underlying
\emph{collective} which is partially hidden to us. If both
phenomena and explanation remains at the same level then nothing
is different from the seventeen century understanding of what
constituted conscious experience (Figure 1). The main difficulty,
and the concern of this proposal, is that there no fundamental
laws yet for the collective of neurons!
\begin{figure}
 \centering \psfig{figure=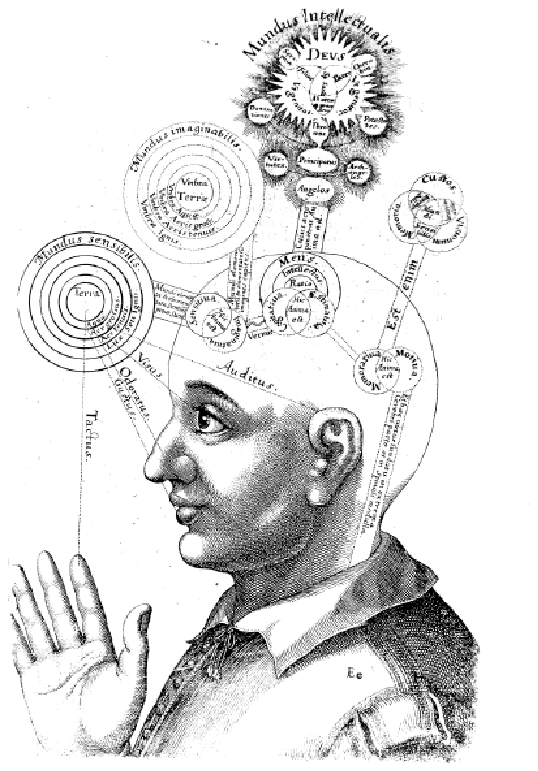,width=3
truein,clip=true,angle=0}
 \caption{Representation of consciousness from the 17th century.
 The macroscopic phenomena, i.e., the imaginary, intellectual
 and sensory world and the respective brain areas remain
 at the same level of description. }
\end{figure}

However, there are some relevant facts which could be source of
inspiration. The brain have, as a collective, some notoriously
conflictive demands. On one side it need to be "integrated" while
must be able to stay "segregated",  as discussed extensively by
Tononi and colleagues \cite{Tononi98,Tononi2004}. This is a non
trivial constraint, nevertheless mastered by the brain as it is
illustrated with plenty of neurobiological phenomenology. Suffice
to think in any conscious experience to immediately realize that
always comprises a single undecomposable scene \cite{Tononi98},
i.e., an integrated state. This integration is such that once a
cognitive event is committed, there is a refractory period (of
about 150 msec.) in which nothing else can be though of. At the
same time the large number of conscious states that can be
accessed over a short time interval exemplify very well the
segregation property. As an analogy, the integration property we
are referring to could be also interpreted as the capacity to act
(and react) on an all-or-nothing mode, similar to an action
potential or a travelling wave in a excitable system. The
segregation property could be then visualized as the capacity to
evoke equal or different all-or-nothing events using different
elements of the system. This could be more than a metaphor.

While the study of this problem is getting increasing attention,
the mechanisms by which this remarkable scenario can exist in the
realms of brain physiology is not being discussed as much as it
should. Our approach is to look at the integration-segregation
dilemma as a generic property of dynamical systems at the critical
point of a phase transition. It is our suggestion that at the
critical point these and others properties -equally crucial for
brain function- appear naturally. If the idea is correct,
statistical physics could help to move the current debate from
phenomenology to understanding of the lower level brain mechanisms
of cognition.
%%%%%%%%%%%%%%%%%%%%%%%%%%%%%%%%%%%%%%%%%%%%%%%%%%%%%%%%%%%

\begin{figure}
\centering \psfig{figure=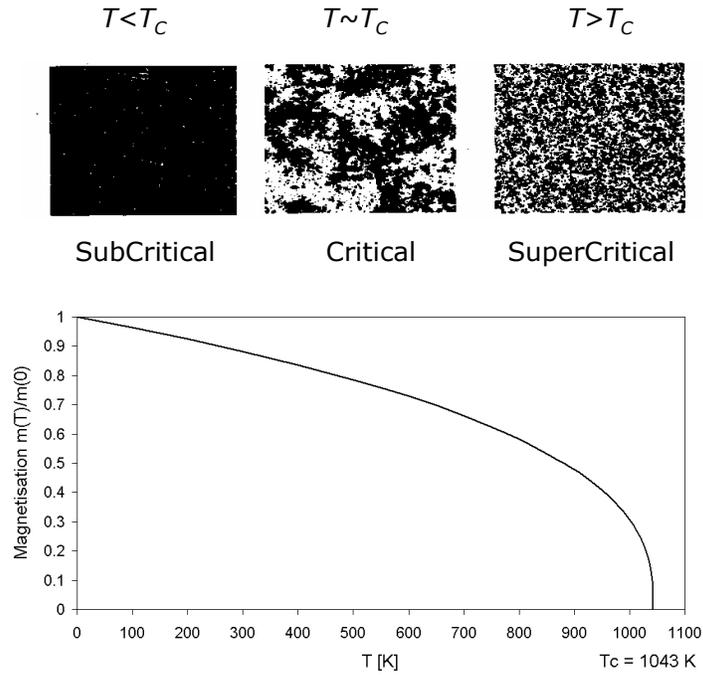,width=4
truein,clip=true,angle=0}

\caption{Ferromagnetic-paramagnetic phase transition. Bottom:
Temperature dependence of magnetization m(T) for Fe. Top three
panels are snapshots of the spins configuration at one moment in
time for three temperatures: subcritical, critical and
supercritical from numerical simulations of the Ising model
(d=2).}
\end{figure}
%%%%%%%%%%%%%%%%%%%%%%%%%%%%%%%%%%%%%%%%%%%%%%%%%%%%%%%%%%%
\subsection{What is special about being critical?} To visualize the
potentially useful connections between brain function and
thermodynamical systems at a phase transition it is helpful to
recall the ferromagnetic-paramagnetic phase transition illustrated
in Figure 2. A material is ferromagnetic if it displays a
spontaneous magnetization in absence of any external magnetic
field. If we heat up an iron magnet the magnetization gets smaller
and finally reaches zero. At low temperature the system is very
ordered with only very large domains of equally oriented spins, a
state that is practically invariant in time. On the other extreme,
at very high temperatures, spins orientation changes constantly,
they are correlated at only very short distances and as
consequently the mean magnetization vanishes. In between these two
homogeneous states, at the critical temperature, the system
exhibits very peculiar fluctuations both in time and space. For
example, the magnetization temporal fluctuations are known to be
scale invariant. Similarly, the spatial distribution of spins
clusters show long range (power law) correlations. At the critical
point, these large dynamic structures emerge, even though there
are only \emph{short-range} interactions between the systems
elements. Thus, at the critical temperature, the system exhibits a
greatly correlated (up to the size of the system) state which at
the same time is able to wildly fluctuate in time at all scales.
We propose that this dynamical scenario -generic for any second
order phase transition- is strikingly similar to the
integrated-segregated dilemma discussed above and shown to be
relevant for the brain to operate as a conscious device. It is
important to note that there is no other conceivable dynamical
scenario or robust attractor known to exhibit these two properties
simultaneously. Of course, any system could trivially achieve
integration and long range correlations in space by increasing
link's strength among faraway sites, but these strong bonds would
prevent any segregated state.

By considering the brain embedded in the rest of nature, one
adopts the Darwinian view that the brains we see today are the
ones that -for whatever means- got an edge and survived. Then we
could ask how consistent is our view of the brain near a critical
point with these Darwinian constraints. We propose that the brains
we see today are critical because the world in which they have to
survive is up to some degree critical as well. If the world were
sub-critical then everything would be simple and uniform (as in
the left panel of Figure 2) there would be nothing to learn, a
brain will be superfluous. In a supercritical world, everything
would be changing all the time (as in the right panel of Figure 2)
it would be impossible to learn. Then we have to conclude that the
brain is only necessary to navigate in a complex, critical world.
In other words we need a brain \emph{because} the world is
critical \cite{bakbook,bak1,bak2,bak3,Mayabak}. Furthermore, a
brain not only have to remember, but also to forget and adapt. In
a sub-critical brain memories would be frozen. In a supercritical
brain, patterns change all the time so no long term memory would
be possible. To be highly susceptible, the brain itself has to be
in the in-between critical state.

These ideas are not knew at all, indeed almost the same intuition
prompted Turing half a century ago to speculate about learning
machines using similar terms:
\begin{quote}
\emph {Let us return for a moment to Lady Lovelace's objection,
which stated that the machine can only do what we tell it to do.
One could say that a man can "inject" an idea into the machine,
and that it will respond to a certain extent and then drop into
quiescence, like a piano string struck by a hammer. Another simile
would be an atomic pile of less than critical size: an injected
idea is to correspond to a neutron entering the pile from without.
Each such neutron will cause a certain disturbance which
eventually dies away. If, however, the size of the pile is
sufficiently increased, tire disturbance caused by such an
incoming neutron will very likely go on and on increasing until
the whole pile is destroyed. Is there a corresponding phenomenon
for minds, and is there one for machines? There does seem to be
one for the human mind. The majority of them seem to be
"subcritical," i.e., to correspond in this analogy to piles of
subcritical size. An idea presented to such a mind will on average
give rise to less than one idea in reply. A smallish proportion
are supercritical. An idea presented to such a mind that may give
rise to a whole "theory" consisting of secondary, tertiary and
more remote ideas. Animals minds seem to be very definitely
subcritical. Adhering to this analogy we ask, "Can a machine be
made to be supercritical?"}
\end{quote}

How things stand today compared with Turing's days? Very
different, because of two important aspects, the first one
concerns with all the advances in monitoring brain signals at
different resolution and the second concerning the possibility to
be guided by the last two decades of results in critical
phenomena.

\subsection{What one should be able to observe?}

A number of features, known to be exhibited by thermodynamic
systems at the critical point, should be immediately observed in
brain experiments, including:

\begin{enumerate}
\item At large scale:\\
 Cortical long range correlations in space and time.\\
 Large scale anti-correlated cortical states.
\item At smaller scale:\\
 "Neuronal avalanches", as the normal homeostatic state for most neocortical
 circuits.\\
 "Cortical-quakes" continuously shaping the large scale synaptic landscape providing "stability" to the cortex.
\item At behavioral level:\\
 All adaptive behavior should be "bursty" and
apparently unstable, always at the "edge of failing".\\
 Life-long learning should be critical due to the effect of continuously "rising the bar".
\end{enumerate}

In addition one should be able to demonstrate that a brain
behaving in a critical world performs optimally at some critical
point, thus confirming the intuition that the problem can be
better understood considering the environment in which brains
evolved.

In the list above, the first item concerns the most elemental
facts about critical phenomena: despite the well known short range
connectivity of the cortical columns, long range structures appear
and disappear continuously. The presence of inhibition as well as
excitation together with elementary stability constraints
determine that cortical dynamics should exhibits large scale
anti-correlated structures as well \cite{Fox2006}. The features at
smaller scales could have been anticipated from theoretical
considerations as well, but avalanches were first observed
empirically in cortical cultures and slices by Plenz and
colleagues \cite{Plenz04}.  An important point that is left to
understand is how these quakes of activity shape the neuronal
synaptic profile during development. At the next level this
proposal suggests that human (and animal \cite{boyer}) behavior
itself should show indications of criticality and learning also
should be included. For example when teaching any skill one
chooses increasing challenge levels which are easy enough to
engage the pupils but difficult enough not to bores them. This
"rising the bar" effect continues trough life, pushing the learner
continuously to the edge of failure! It would be interesting to
measure some order parameter for sport performance to see if shows
some of these features for the most efficient teaching strategies.

%%%%%%%%%%%%%%%%%%%%%%%%%%%%%%%%%%%%%%%%%%%%%%%%%%%%%%%%%%%%%%%%%%
\begin{figure}
\centering \psfig{figure=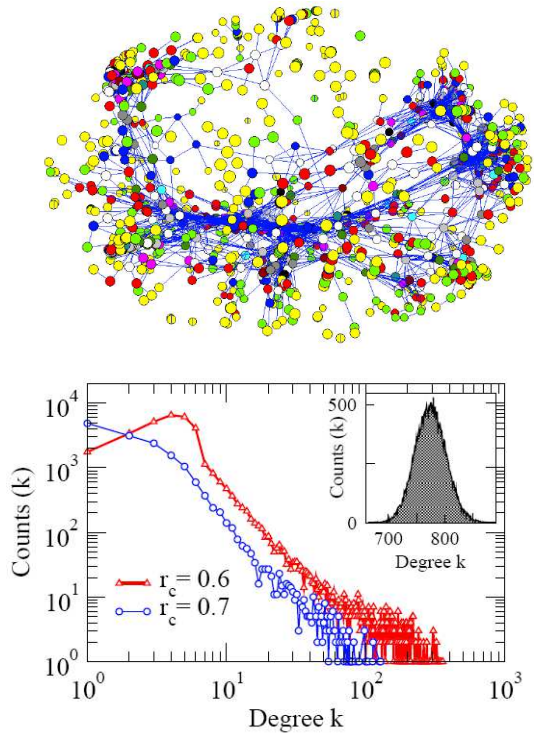,width=3.0
truein,clip=true,angle=0} \caption{A typical brain network
extracted from functional magnetic resonance imaging. Top panel
shows a pictorial representation of the network. The bottom panel
shows the degree distribution for two correlation thresholds
$r_c$. The inset depicts the degree distribution for an equivalent
randomly connected network. Data re-plotted from \cite{Eguiluz}.}
\end{figure}
%%%%%%%%%%%%%%%%%%%%%%%%%%%%%%%%%%%%%%%%%%%%%%%%%%%%%%%%%%%%%%%%%
\section{Recent results}
\subsection{Functional brain networks are complex}
Functional magnetic resonance imaging (fMRI) allows to monitor non
invasively spatio-temporal brain activity  under various cognitive
conditions. Recent work using this imaging technique demonstrated
complex functional networks of correlated dynamics responding to
the traffic between regions, during behavior or even at rest (see
methods in \cite{Eguiluz}. The data is analyzed in the context of
the current understanding of complex networks (for a review see
\cite{Sporns2004}). During any given task the networks are
constructed first by calculating linear correlations between the
time series of brain activity in each of $36\times64\times64$
brain sites. After that, links are said to exist between those
brain sites whose temporal evolutions are correlated beyond a
pre-established value $r_c$.

Figure ~2, show a typical brain functional network extracted with
this technique. The top panel illustrates the interconnected
network's nodes and the bottom panel shows the statistics of the
number of links (i.e., the degree) per node.   There is a few very
well connected nodes in one extreme and a great number of nodes
with a single connection.  The typical degree distribution
approaches a power law distribution with an exponent around 2.
Other measures revealed that the number of links as a function of
-physical- distance between brain sites also decays as a power
law, something already confirmed by others \cite{Salvador} using
different techniques. Two statistical properties of these
networks, path length and clustering were computed as well. The
path length ($L$) between two voxels is the minimum number of
links necessary to connect both voxels. Clustering ($C$) is the
fraction of connections between the topological neighbors of a
voxel with respect to the maximum possible.  Measurements of $L$
and $C$ were also made in a randomized version of the brain
network. $L$ remained relatively constant in both cases while $C$
in the random case resulted much smaller, implying that brain
networks are "small world" nets, a property with several
implications in terms of cortical connectivity, as discussed
further in \cite{Sporns2004b,Sporns2004}. In summary, the work in
\cite{Eguiluz} showed that functional brain networks exhibit
highly inhomogeneous scale free functional connectivity with small
world properties. Although these results admit a few other
interpretations, the long range correlations demonstrated in these
experiments are consistent with the picture of the brain operating
near a critical point. Of course, further experiments are needed
to specifically define and measure some order parameter to clarify
the precise nature of these correlations. Furthermore, as more
detailed knowledge of the properties of these networks is
achieved, the need to integrate this data in a cohesive picture
grows as discussed recently by Sporns and colleagues
\cite{Sporns2006}.
%%%%%%%%%%%%%%%%%%%%%%%%%%%%%%%%%%%%%%%%%%%%%%%%%%%%%%%%%%%%%%%%%
\begin{figure}
\centering \psfig{figure=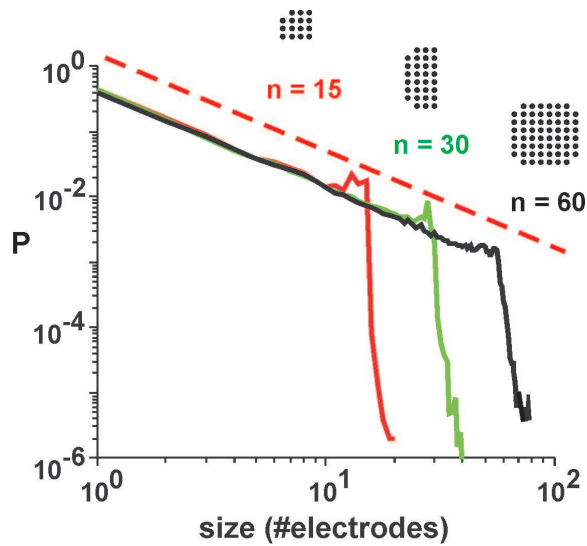,width=3
truein,clip=true,angle=0}

\caption{The size distribution of neuronal avalanches in mature
cortical cultured networks follows a power law with an exponent
$\sim 3/2$ (dashed line). The data, re-plotted from Figure 4 of
\cite{Plenz04}, shows the probability of observing an avalanche
covering a given number of electrodes for three sets of grid sizes
shown in the insets with n=15, 30 or 60 sensing electrodes
(equally spaced at $200 \mu m$). The statistics is taken from data
collected from 7 cultures in recordings lasting a total of 70
hours and accumulating 58000 ($+-$ 55000) avalanches per hour
(mean $+-$ SD).}
\end{figure}
%%%%%%%%%%%%%%%%%%%%%%%%%%%%%%%%%%%%%%%%%%%%%%%%%%%%%%%%%%%%%%%
\subsection{Cortical networks exhibit neuronal avalanches} Recent experiments from Plenz and
colleagues \cite{Plenz04} were the first to demonstrate a new type
of small scale cortical activity. They showed that under some
experimental conditions, the cortex exhibits what they termed
neuronal avalanches. This type of population activity seats half
way in between two well known patterns: the oscillatory or
wave-like highly coherent activity on one side and the asynchronic
and uncoherent modality on the other. In each avalanche neuronal
activity have a very large probability to engage few neurons and
die, and a very low probability to spread and activate the whole
system. In very elegant experiments Plenz and colleagues estimated
a number of properties indicatives of critical behavior including
a power law with an exponent $\sim 3/2$ for the density of
avalanche sizes (see Figure 4). This agrees exactly with the
theoretical expectation for a critical branching process
\cite{Zapperi1995}. Further experiments in other experimental
settings, including monkey and rats in vivo recordings, have
already confirmed and expanded these initial estimations
\cite{Plenzinpress, Stewart2006}. An unsolved problem here is to
elucidate the precise neuronal mechanisms leading to this
behavior. Avalanches of activity such as the one observed by Plenz
and colleagues could be the reflection of completely different
scenarios. It could reflects a structural (i.e., anatomical)
substrate over which travelling waves in the peculiar form of
avalanches occur. This will imply that the long range correlations
detected are trivially due to long range connections. If that is
the case, as was discussed above, this have nothing to do with
criticality, and furthermore it will imply that segregation will
be impossible. The second possibility is that avalanches occurs
over a population of locally connected neurons. Their ongoing
collective history will permanently keep them near the border of
avalanching and each avalanche will only excite enough neurons to
dissipate the excess of activity. Although this is the most likely
scenario, which follows the ideas and results put forward by Bak
and colleagues
 \cite{bakbook,bak1,bak2,bak3,Mayabak}, there is much theoretical
work awaiting to formalize these results.

\begin{figure}

\centering \psfig{figure=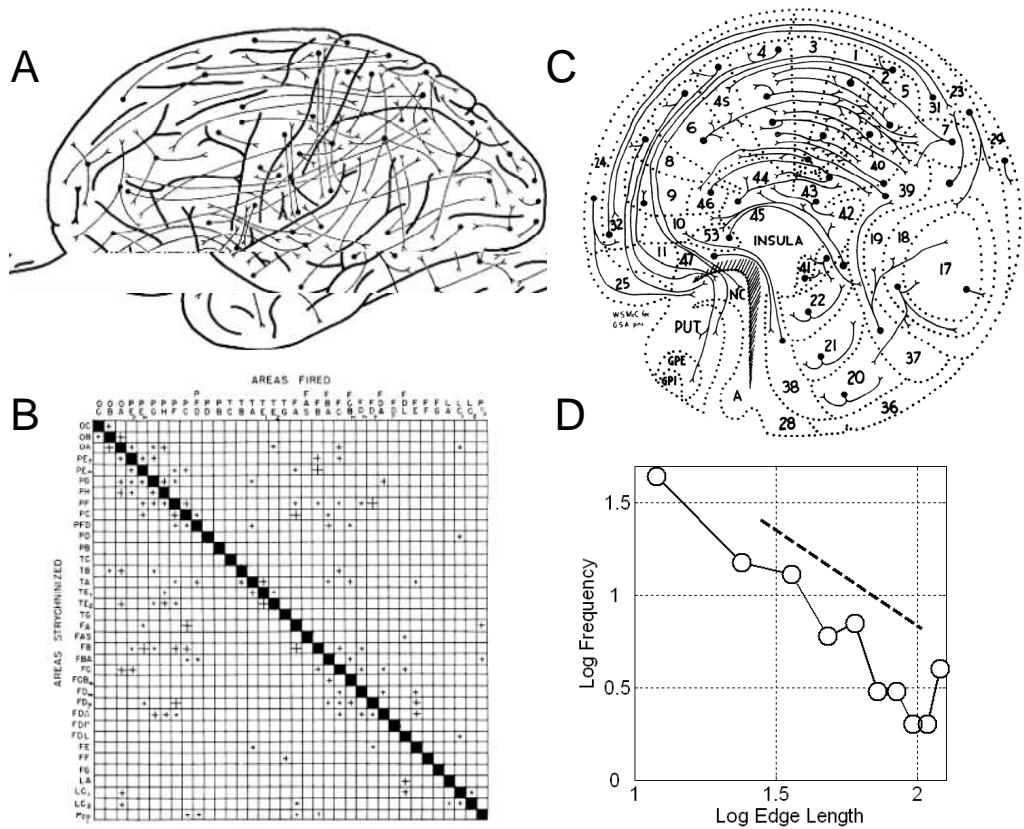,width=6
truein,clip=true,angle=0}

 \caption{McCulloch experiments
inducing local seizures by instillation of strychnine. Panel A and
B are from chimpanzee experiments \cite{McCullochbook}. Panel A
shows a summary of the sites where the strychnine was applied
(filled circled) and the sites of the cortex fired  by the topical
application. Besides the local ones, long range activations
crossing the entire cortex were often observed. Panel B
illustrates the adjacency matrix summarizing which areas -on the
average- were activated by the strychnine application. Panel C
shows similar results obtained by McCulloch and colleagues in
Macaca Mulata \cite{McCulloch1944} mapping the entire cortex and
basal ganglia. Panel D depicts (note the double logarithmic axis)
the edge length density distribution computed from McCulloch's
drawing in Panel A. The dashed line with slope 2 illustrates, for
comparison, the average edge-length density found in recent fMRI
experiments \cite{Eguiluz}.}
\end{figure}

\subsection{McCulloch already saw it in 1940}

Dusser de Barenne, Warren McCulloch and colleagues
\cite{McCulloch1944,McCulloch1941}, more than sixty years ago,
experimented inducing local seizures by instillating drops of
strychnine in several regions of the monkey cortex while recording
cortical electrical activity simultaneously in twenty sites across
the entire cortex. This clever technique, mastered by Dusser de
Barenne,  received the name of strychnine neuronography. In a
certain way, these experiments could be considered the earliest
attempt to study brain functional connectivity, by inducing some
liminal activity in a given area and recording the co-active
cortical sites. Typically, they noticed that the initial activity
induced by the strychnine remained local, and did not generalized
to the entire cortex. However with surprise they noted that, less
often, the activity was recorded in very far away locations.
Figure 5 (redrawn from the original sketches in
\cite{McCullochbook}) summarizes these early observations together
with our own rough estimations in Panel D. Filled circles in Panel
D represent the distribution of edge lengths, computed from the
drawing in Panel A  as the Euclidean distance (using arbitrary
units) between the location of each strychnine instillation and
the resulting activation site/s. Note that, despite the scarcity
of the data, the results demonstrate long range correlations, the
exponent being similar to the estimations using fMRI
\cite{Eguiluz,Salvador}. For example an application in the frontal
cortex induced activity sometimes in the occipital cortex.
Nowadays, is not difficult to admit that frontal activation will
evoke visual imaginery and viceversa, however McCulloch knew that
much before us.

\subsection{Senses are critical}
In more than one sense our senses seems to be critical. To move
around, to escape from predators, to choose a mate or to find
food, the sensory apparatus is critical for any animal survival.
But it seems that senses are also critical in the thermodynamic
sense of the world. Consider first the fact that the density
distribution of the various form of energy around us is clearly
inhomogeneous, at any level of biological reality, from the sound
loudness any animal have to adapt to the amount of rain a vegetal
have to take advantage. From the extreme darkness of  a deep cave
to the brightest flash of light there are several order of
magnitude changes, nevertheless our sensory apparatus is able to
inform the brain of such changes. It is well known that isolated
neurons are unable to do that because of their limited dynamic
range, which spans only a single order of magnitude. This is the
oldest unsolved problem in the field of psychophysics, tackled
very recently by Kinouchi and Copelli \cite{kinouchi} by showing
that the dynamics emerging from the \emph{interaction of coupled
excitable elements} is the key to solve the problem. Their results
show that a network of excitable elements set precisely at the
edge of a phase transition - or, at criticality - can be both,
extremely sensitive to small perturbations and still able to
detect large inputs without saturation. This is generic for any
networks regardless of the neurons' individual sophistication. The
key aspect in the model is a local parameter that control the
amplification of any initial firing activity. Whenever the average
amplification is very small activity dies out; the model is
subcritical and not sensitive to small inputs. On the other hand,
choosing an amplification very large one sets up the conditions
for a supercritical reaction in which for any - even very small -
inputs the entire network fires. It is only in between these two
extremes that the networks have the largest dynamic range. Thus,
amplification around unity, i.e., at criticality, seems to be the
optimum condition for detecting large energy changes as an animal
encounters in the real world \cite{chialvo2006}. It is only in a
critical world that energy is dissipated as a fractal in space and
time with the characteristic highly inhomogeneous fluctuations.
Since the world around us appears to be critical, it seems that
we, as evolving organisms embedded in it, have no better choice
than to be the same.

\section{Outlook}
The preceding section purposely presented only a selection of
concrete results inspired in the approach promoted here. They do
not probe that the brain is critical, but they demonstrate that
there are relevant aspects of brain dynamics which underlying
collective is critical in some sense. There are, of course, an
increasingly large body of work modelling and explaining further
these experimental findings, which we will not enumerate, because
this is not an exhaustive review. An excellent survey is in press
and we direct the readers to it
 \cite{PlenzTINS}. Nevertheless we mention, mostly as a guide for
further reading, ideas connected with the general framework
discussed here. Probably the first to note should be Ashby's work
to understand how the forces of self-organization could shape a
brain \cite{Ashby}. The work of Tononi, Edelman and colleagues
\cite{Tononi98a, Tononi98} it is the first to delineate the
fundamental problem of integration and segregation and to explore
its connection with complexity. The analysis of cortical
coordination dynamics discussed by Kelso, Bressler and colleagues
\cite{Kelso}, are related with this proposal, because it main
ingredients, collective variables, and metastable coordination
states are all generic of the critical state discussed here. Of
note also is Dehaene \cite{Dehaene2001} "workspace" model of
conscious experience that resemble the scale free distribution of
hubs observed experimentally and discussed above. Most probably a
detailed analysis of their specific numerical models would reveal
optimum performance near criticality, something worth to pursue.
Finally, there is the exhaustive review of Werner
\cite{Werner2006} advocating to further the study of phase
transitions, metastability and criticality in cognitive models and
experiments.

The main difference that set apart this proposal from all of the
above efforts, is that it does not pretend to be novel or ad hoc.
Right or wrong, but deliberately, the proposal is that relevant
aspects of brain dynamics \emph{can} be understood using the same
theoretical framework as for any nonequilibrium thermodynamic
system at or near the critical point of a second order phase
transition.

Arguably, brain theory is still at a stage comparable to physics
in "pre-thermodynamic" times. Imagine yourself in days previous to
the notion of temperature. Similarities between scalding water and
ice will be supported by their similar "burning" (to the touch)
properties, when hot or cold were only subjective quantities. Of
course, the notion of pressure and temperature together with
phases changed everything. Brain theory will eventually undergo
such transformation  starting with the preliminary definition of
order parameters such as Tononi's $\Phi$ \cite{Tononi2004} and the
elaboration of some phase diagram, including degrees of
consciousness, modalities of transitions between phases, etc.
Until then, pre-thermodynamic debates will surely continue.

\begin{theacknowledgments}
Work supported by NIH NINDS of USA (Grants 42660 and 35115). The
warm hospitality of the colleagues of Universidad de Granada are
also acknowledged. Special thanks to Dr. Dietmar Plenz (NIMH) for
stimulating discussions and for providing Fig. 4.
\end{theacknowledgments}

%\endinput
%%
\end{document}